\documentclass[letterpaper]{emulateapj}
\usepackage{amsmath}
\usepackage{graphics}
\usepackage{amsmath,cases}
\usepackage{color}
\usepackage{ulem}
\usepackage{lineno}

\usepackage{mathrsfs}
\newif\ifoneauthor
\oneauthortrue
\newcommand{\unit}[1]{\ {\rm #1}}

\newcommand{\Eq}[1]{Eq. (\ref{#1})}
\newcommand{\Table}[1]{Table \ref{#1}}
\newcommand{\Fig}[1]{Figure \ref{#1}}

\DeclareMathAlphabet{\mathpzc}{OT1}{pzc}{m}{it}
\usepackage{here}
\usepackage{float}
\definecolor{gray}{gray}{0.4}

\begin{document}

\title{Probe for Type Ia supernova progenitor in decihertz gravitational wave astronomy}



\author{Tomoya Kinugawa}
\email[]{kinugawa@icrr.u-tokyo.ac.jp}
\affiliation{Institute for Cosmic Ray Research, University of Tokyo, Kashiwa, Chiba 255-8582, Japan}
\author{Hiroki Takeda}
\email[]{takeda@tap.scphys.kyoto-u.ac.jp}
\affiliation{Department of Physics, Kyoto University, Kyoto 606-8502, Japan}
\affiliation{Department of Physics, University of Tokyo, Bunkyo, Tokyo 113-0033, Japan}
\author{Ataru Tanikawa}
\email[]{tanikawa@ea.c.u-tokyo.ac.jp}
\affiliation{Department of Earth Science and Astronomy, College of Arts ans Sciences, The University of Tokyo, 3-8-1 Komaba, Meguro-ku, Tokyo 153-8902, Japan}
\author{Hiroya Yamaguchi}
\email[]{yamaguchi@astro.isas.jaxa.jp}
\affiliation{Institute of Space and Astronautical Science, JAXA, 3-1-1 Yoshinodai, Sagamihara, Kanagawa 229-8510, Japan}
\affiliation{Department of Physics, University of Tokyo, Bunkyo, Tokyo 113-0033, Japan}



\date{\today}

\begin{abstract}
It is generally believed that Type Ia supernovae are thermonuclear explosions of carbon-oxygen white dwarfs  (WDs).
However, there is currently no consensus regarding the events leading to the explosion.
A binary WD (WD-WD) merger is a possible progenitor of Type Ia supernovae. 
Space-based gravitational wave (GW) detectors with considerable sensitivity in the deci-Hz range such as the DECi-hertz Interferometer Gravitational wave Observatory (DECIGO) can observe WD-WD mergers directly.
Therefore, access to the deci-Hz band of GWs would enable multi-messenger observations of Type Ia supernovae to determine their progenitor and explosion mechanism. 
In this paper, we consider the event rate of WD-WD mergers and minimum detection range to observe one WD-WD merger per year, using {a} nearby galaxy catalog and the relation between the Ia supernova and host galaxy. Furthermore, we calculate DECIGO's ability to localize WD-WD mergers and to determine the masses of binary mergers.
We estimate that the deci-Hz GW observatory can detect GWs with amplitudes $h\sim10^{-20}[\rm Hz^{-1/2}]$ at {0.01-0.1 Hz}, which is 1000 times higher than the detection limit of DECIGO.
   {Assuming progenitors of Ia supernovae are merging WD-WD ($1M_{\odot}-0.8M_{\odot}$), DECIGO is expected to detect 6600 WD-WD mergers within $z=0.08$, 
  and identify the host galaxy of such WD-WD mergers within $z\sim0.065$ using GW detection alone.}
\end{abstract}


\maketitle

\section{Introduction}\label{intro}
Advanced LIGO have detected gravitational waves (GWs) from compact binary mergers. These
GWs have reveal the existence of massive stellar black holes, and the
origin of the r-process elements and short gamma-ray burst
\citep{Abbot2016a,Abbot2016b,Abbott2017}.  We are at the dawn of
GW astronomy.  Ground GW detectors such as advanced LIGO, advanced
VIRGO, and KAGRA cover 10--10,000\,Hz. 
{LISA is designed to detect 
milli Hz GWs and will launch during the 2030s \citep{LISA2017,LISA2022}. The design sensitivity of LISA is 
$h\sim10^{-20}$$[\rm Hz^{-1/2}]$ at 0.01 Hz}. DECIGO and
B-DECIGO, a test version of DECIGO, fill the gap between the ground GW
detectors and LISA \citep{Seto2001, Nakamura2016}.  The main target of
DECIGO is the stochastic background GW from the early universe.
Design sensitivity of DECIGO and B-DECIGO are
$h\sim10^{-23}$-$10^{-24} [\rm Hz^{-1/2}]$, and
$h\sim10^{-22}$-$10^{-23} [\rm Hz^{-1/2}]$ at 0.1--10 Hz, respectively.
These sensitivities are very useful to detect high redshift binary
black hole mergers and check the origin of massive stellar binary
black holes
\citep[e.g.][]{Kinugawa2014,Kinugawa2016,Belczynski2016,Nakamura2016,Kinugawa2020,Tanikawa2022}.
Furthermore, the GW from a white dwarf-white dwarf (WD-WD) merger is also {$\sim 0.01-0.1$ Hz
\citep{Dan2011,Mandel2018}}.  Thus, WD-WD mergers will become interesting
science targets of DECIGO \citep{Seto2001,Mandel2018} and other 0.1 Hz
GW detectors such as B-DECIGO \citep{Nakamura2016}, TianGO
\citep{TianGO2019}, DO\citep{DO2019} and AMIGO
\citep{AMIGO2019}. 

WD-WD mergers are one of the most promising candidates of Ia supernova
progenitors.
Although Ia supernovae have considerable significance as distance
indicators in cosmology, many fundamental aspects of their evolution
and explosion are still under debate {\citep[see a recent review by][]{Ruiter2020}. It is widely accepted that the Ia supernova progenitor system is a carbon-oxygen (C-O) WD in an interacting binary star. However, the nature of the companion star and the WD mass are unclear. {The companion star can be a non-degenerate star (main sequence, giant-like or stripped-helium-burning star)}, or another WD, also known as single degenerate (SD) and double degenerate (DD) scenarios, respectively. The WD mass can be near the Chandrasekhar limit, $\sim 1.4M_{\odot}$, (Chandrasekhar-mass explosion), or below the limit (sub-Chandrasekhar-mass explosion). For the SD scenario with the Chandrasekhar-mass explosion, the WD approaches the Chandrasekhar limit via mass transfer from its non-degenerate companion star \citep[e.g][]{Whelan1973, Nomoto1982a, Hachisu1996}. For the SD scenario with the sub-Chandrasekhar-mass explosion, a C-O WD accretes helium-rich materials from its companion helium star, and initiates double detonation explosion in which helium detonation on the WD surface {leads to detonation of carbon} in the WD core \citep{Nomoto1982b, Woosley1986, Livne1990}.}

{The DD scenario with Chandrasekhar-mass explosion can occur tidal disruption of the lighter WD followed by thermal mass accretion onto the heavier WD \citep[e.g.][]{Webbink1984,Iben1984}. 
For the DD scenario with sub-Chandrasekhar-mass explosion, three things can occur. First, a dynamical merger of two C-O WDs initiates carbon detonation directly, referred to as a carbon-ignited violent merger \citep[e.g.][]{Pakmor2010,Sato2015}. Second, it can result in another explosion mode known as spiral instability \citep{Kashyap2015, Kashyap2017}. Lastly, a dynamical mass accretion onto the heavier C-O WD from lighter helium-rich WD triggers double detonation explosion, referred to as a helium-ignited violent merger or a dynamically-driven double-degenerate double-detonation (D6) \citep[e.g.][]{Guillochon2010, Fink2010, Woosley2011, Pakmor2013, Shen2014}.
{In the DD scneario with Chandrasekhar explosion, Ia supernovae may occur $\sim100$ yr or more after the WD-WD mergers due to slow mass accretion.} Conversely, in the DD scenario with sub-Chandrasekhar-mass explosion, Ia supernovae {occur promptly around the time of WD-WD merger}. Thus, if any sub-Chandrasekhar-mass explosion models {do occur}, WD-WD mergers emit GW and electromagnetic (EM) signals at the same time, it can be a promising target of multi-messenger astronomy. This can be applied to identifying progenitors of subclasses of Ia supernovae.}

In this paper, we consider the event rate of WD-WD mergers and minimum detection range 
to observe one WD-WD merger per year, using the nearby galaxy catalog and the relation 
between the Ia supernova and host galaxy (\S \ref{s.rate}). Furthermore, we investigate the observational performance of 
DECIGO for WD-WDs by the parameter estimation of the inspiral GWs from WD-WD mergers.
Mainly, we eevaluate the detection rate and the ability of these detectors to localize WD-WD mergers and to determine the masses of binary mergers (\S \ref{inspiral}, \ref{PE}, and \ref{Results}).
Throughout this paper, we use CGS units except for \S \ref{inspiral}.


\section{Event rate of WD-WD mergers}
\label{s.rate}
In order to estimate the WD-WD merger rate, we assume that all the Ia supernova progenitors are WD-WD mergers.
{Note that this assumption ignores the possibility of the SD case. However, \cite{Maoz2018} shows that the expected WD-WD merger rate in the Milky Way is  4.5--7 times more than its specific Ia supernova rate. Thus, the WD-WD merger rate from our assumption might be smaller than the actual value.}

The Ia supernova rate at $z\sim0$ is 
\begin{equation}
(0.301 \pm0.062)\times10^{-4}~\rm SN~yr^{-1}~Mpc^{-3},
\label{rate}
\end{equation}
determined from the Lick Observatory Supernova
Search (LOSS) \citep{Li2011}.
We use this value as the fiducial rate of WD-WD mergers to estimate the detection rate in this paper.
However, this rate is the averaged volumetric rate of Ia supernova.
 {The relation between WD-WD mergers and Ia supernovae can be confirmed by the early detection of the first GW from a WD-WD merger and its electromagnetic (EM) counterpart. This is similar to the GW170817 where the relation between binary neutron star mergers and short GRBs+kilonovae was revealed  \citep{Abbott2017}.
We need to determine the minimal detection volume to detect a WD-WD merger per year in the nearby galaxies.}
{There are many studies on {the rate-size relation of Ia supernova host galaxies} \cite[e.g.][]{Sullivan2006,Totani2008,Li2011,Childress2014,Graur2015} and these results indicate the relation between Ia supernova rate and the stellar mass of the galaxies.}
{In order to estimate the event rate of Ia supernova near our galaxy, we consider the rate-size relation of Ia supernova host galaxy \citep{Li2011} and the catalog of nearby galaxies within 11 Mpc \citep{catalog2013}.}
The rate-size relation is 
\begin{equation}\label{SNuM}
    {\rm SNuM=SNuM}(M_0)\times\left(\frac{M_*}{10^{10}M_{\odot}}\right)^{\rm RSS},
\end{equation}
where SNuM, SNuM($M_0$), $M_*$, and RSS are the Ia supernova rate per century per $10^{10}M_{\odot}$, the normalization value, the stellar mass of the galaxy, and the power law index,  respectively. 
\cite{Li2011} obtained $-0.513\pm 0.316,~-0.503\pm0.158,~-0.637\pm0.199,~-0.555\pm0.171,~-0.443\pm0.241,~-0.329\pm0.201$, and $-0.435\pm0.195$ as RSSs for E, S0, Sa, Sb, Sbc, Sc, and Scd galaxies, respectively. 
The combined significance with RSS = -0.5 and SNuM($M_0$) = 0.25 for E, S0, Sa, Sb, Sbc, Sc, and Scd galaxies is $-7.4\sigma$ \citep{Li2011}.
Thus, we adopt $-0.50$ and 0.25 as RSS and SNuM$(M_0)$ in our rate calculations, respectively. 

\begin{figure}[!ht]
	\begin{center}
		\includegraphics[width=1.0\hsize]{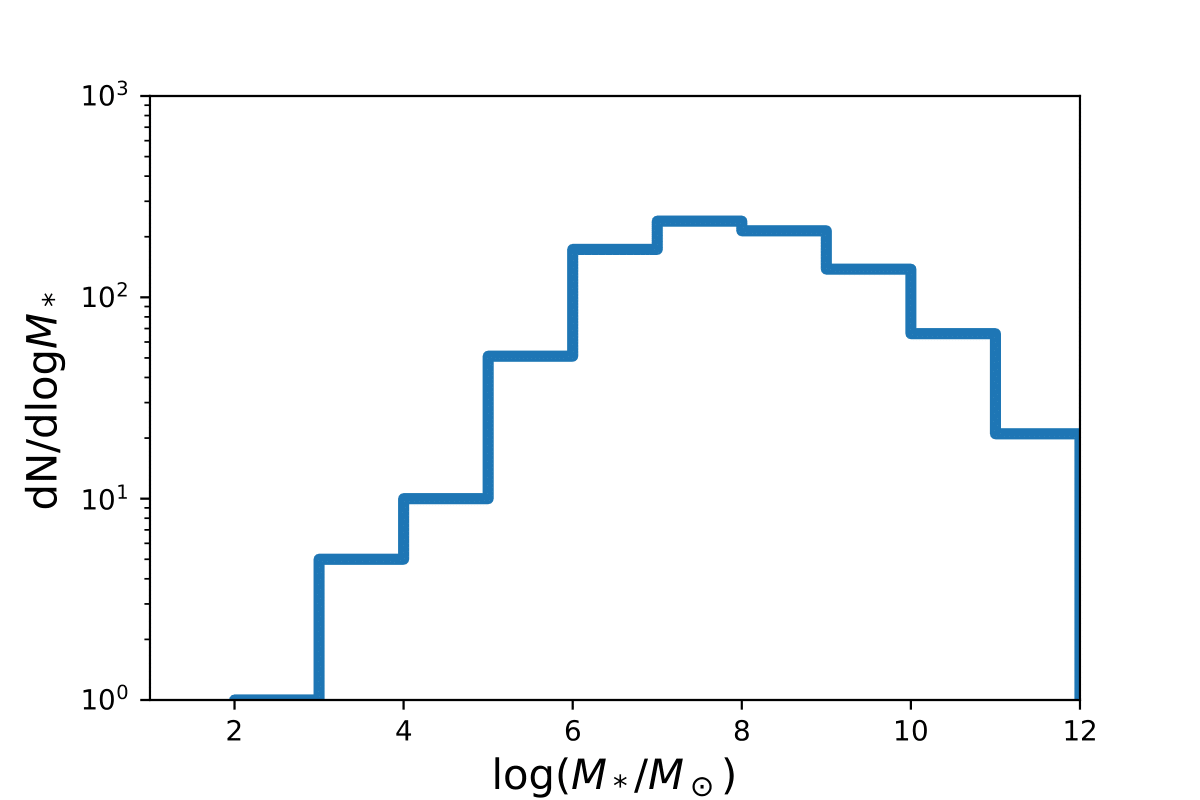}
	\end{center}
	\caption{Stellar mass distribution of nearby galaxies within 11 Mpc.
		\label{Gmass}}
\end{figure}

We use the galaxy catalog of \cite{catalog2013} in order to get the stellar mass of nearby galaxies.
In this catalog, there are data on 1209 galaxies within 11 Mpc and 951 galaxies' masses or lower mass limits.
Figure \ref{Gmass} shows the stellar mass distribution of nearby galaxies.
In the case of galaxies whose masses are $\lesssim10^{5}\rm~M_{\odot}$, we use lower mass limits of galaxies as galaxy stellar masses.

{We calculate the Ia supernova rate for each nearby galaxy using Eq.\ref{SNuM} and stellar masses of nearby galaxies in this catalog data, and then take the sum of the Ia supernova rate in nearby galaxies within 11 Mpc.
The summation of the Ia supernova rate in nearby galaxies is
\begin{equation}
\label{11Mpc_rate}
    0.85~\rm~yr^{-1}.
\end{equation}
This rate shows that approximately one Ia supernova will occur about once a year in nearby galaxies.
If the detection range of GW detector on $\sim$0.01--0.1 Hz band is more than 11 Mpc, the combination with the EM observation of supernova and the GW observation can be used to determine whether Ia supernova progenitors are WD-WD mergers or not.
}

\section{Gravitational waveform for inspiraling binary white dwarf}
\label{inspiral}
In this section, we use $c=G=1$.
In general relativity, the detector signal of a GW from an inspiraling binary system in time domain can be written as \citep{Berti2005, Maggiore2007},
\begin{equation}
\label{signal_time}
h(t)\simeq\frac{2m_{1}m_{2}}{r_s(t)d_L}\mathscr{A}(t)\cos{(\int^tf_{\rm gw}(t')dt'+\phi_p(t)+\phi_D(t))},
\end{equation}
where $m_1, m_2$ are the masses of binary stars, $r_s(t)$ is the orbital relative distance, $d_L$ is the luminosity distance to the binary system, $f_{\rm gw}$ is the frequency of the GW, and $\phi_D(t)$  is the Doppler phase. $\mathscr{A}(t)$ and $\phi_p(t)$ are defined by,
\begin{equation}
\mathscr{A}(t):=\sqrt{(1+\cos^2{\iota})^2F^{+}(t)^2+4\cos^2{\iota}F^{\times}(t)^2},
\end{equation}
\begin{equation}
\phi_p(t):=\arctan\left(\frac{2\cos{\iota}F^{\times}(t)}{(1+\cos^2{\iota})F^{+}(t)}\right).
\end{equation}
where $\iota$ is the inclination angle of the binary system. Here, $F^{+}(t)$ is the antenna pattern functions for the plus polarization mode, and $F^{\times}(t)$ is that for the cross mode \citep{Nishizawa2009}.

As $(2m_{1}m_{2})/({r_s(t)d_L}), \mathscr{A}(t), \phi_p(t), \phi_D(t)$ vary in time slowly compared to $\int f_{\rm gw}(t')dt'$, we can calculate the Fourier component $\tilde{h}(f)$ of the detector signal $h(t)$ by the stationary phase approximation \citep{Maggiore2007, Berti2005, Cutler1998, Arun2006, Takeda2019},
\begin{equation}
\label{signal_tensor}
 \tilde{h}(f)=\mathcal{A}f^{-7/6}e^{i\Psi(f)}\mathcal{G}_{T}(t(f)).
\end{equation}
with the GW amplitude $\mathcal{A}$ and the phase $\Psi(f)$.
The geometrical factor for tensor modes $\mathcal{G}_{T}$ is defined by
  \begin{eqnarray}
 \label{geo_tensor}
 \mathcal{G}_{T}(t):=\frac{5}{4}\mathscr{A}(t)
 e^{i(\phi_p(t)+ \phi_{D}(t))}.
 \end{eqnarray}
 $t(f)$ gives the relation between the time to coalescence and the frequency of the GW before merger \citep{Maggiore2007, Damour2001, Damour2002}, which is defined by the condition $f=f_{\rm gw}(t(f))$,
 \begin{equation}
 \label{tf}
 t(f):=t_c-\frac{5}{256}\mathcal{M}^{-5/3}(\pi f)^{-8/3},
 \end{equation} 
where $\mathcal{M}:=(m_1m_2)^{3/5}(m_1+m_2)^{-1/5}$ is the chirp mass and $t_c$  is the coalescence time. 

{We adopt the inspiral waveform up to Newtonian order in amplitude $\mathcal{A}$  and 2.5 post-Newtonian (PN) order in phase $\Psi(f)$ \citep{Santamaria2010, Khan2016},}
\begin{equation}
\label{amp}
\mathcal{A}f^{-7/6}=\frac{1}{\sqrt{30}\pi^{2/3}d_L}\mathcal{M}^{5/6}f^{-7/6},
\end{equation}
and
\begin{equation}
\label{phase}
\Psi(f)=2\pi ft_c-\phi_c-\frac{\pi}{4}+\frac{3}{128}(\pi\mathcal{M}f)^{-5/3}\sum_{i=0}^{5}\phi_i(\pi\mathcal{M}f)^{i/3}.
\end{equation}
where $\phi_c$ is the phase at the coalescence time and $\phi_i$ are PN coefficients.
 The amplitude is kept up to the Newtonian order because of the consistency with the order in \Eq{tf}. The binary eccentricity is not considered for simplicity because WD-WDs are expected to have circular orbits due to tidal interactions \citep{Willems2007,Ruiter2010}. It has been reported that the deformations due to filling the Roche lobe or the existence accretion disk induce typical difference at the level of one percent or less for semi-detached WD-WDs with respect to the average strain amplitude \citep{Broek2012}. For the WD-WDs whose rotations are synchronized with the orbital motion, it has been reported that finite size effects and certain universal relation are helpful to identify the individual masses of binary WDs \citep{Wolz2021}. 
 {Effects of mass transferring and the tidal effect just before mergers are studied by \cite{Kremer2017} and \cite{McNeill2020}, respectively.
}
However, the chirp effect is larger in the decihertz band because WD-WDs can be observed until just before the merger. Therefore, we concentrate on the mass and position information that can be extracted from the inspiral chirp signal alone, and ignore these effects of the WDs for the sake of generality.

\section{Parameter estimation}
\label{PE}
  In order to investigate the possibility of identifying the properties of the WD-WDs as a progenitor of Ia supernova in decihertz GW astronomy, we conducted parameter estimation by Fisher analysis \citep{Finn1992, Cutler1994}. A Fisher information matrix gives the  Cramer--Rao bound of the system parameter. In other words, a Fisher information matrix tells us how precisely we can determine the model parameters by observations under strong signal and Gaussian noise assumptions. 
The Fisher information matrix $\Gamma$ is calculated by
\begin{equation}
\Gamma_{ij}:=4{\rm{Re}}\int^{\rm{f_{max}}}_{\rm{f_{min}}}df\sum_I \frac{1}{S_{n,I}(f)}\frac{\partial \tilde{h}^{*}_I(f)}{\partial\lambda^i}\frac{\partial \tilde{h}_I(f)}{\partial\lambda^j},
\label{Fisher}
\end{equation} 
 where $S_{n,I}(f)$ is the I-th detector noise power spectrum and $\lambda^i$ is the i-th binary parameter. 
The inverse of the Fisher information matrix gives the root mean square error of a parameter $\Delta\lambda^i$, calculated by 
\begin{equation}
(\Delta\lambda^i)_{\rm rms}:=\sqrt{\langle\Delta\lambda^i\Delta\lambda^i\rangle}=\sqrt{(\Gamma^{-1})^{ii}},
 \end{equation}
 where $\Delta\lambda^i$ is the measurement error of $\lambda^i$ and $\langle\cdot\rangle$ denotes ensemble average. Then, the sky localization error is defined by
 \begin{equation}
 \Delta\Omega_s:=2\pi|\sin{\theta_s}|\sqrt{\langle(\Delta\theta_s)^2\rangle\langle(\Delta\phi_s)^2\rangle-\langle\Delta\theta_s\Delta\phi_s\rangle^2}.
 \end{equation}
 Hereafter, we simply refer to $(\Delta\lambda_i)_{\rm rms}$ as $\Delta\lambda_i$, and call it the estimation error of $\lambda_i$.

we have 11 model parameters in GR
 \begin{equation}
(\log\mathcal{M},\log{\eta}, t_c, \phi_c, \log{d_L}, \chi_s, \chi_a, \theta_s, \phi_s, \cos{\iota}, \psi_p),
\end{equation}
where $\log{\eta}, \chi_s, \chi_a$ are the logarithm of the symmetric mass ratio $\eta:=m_1m_2/(m_1+m_2)^2$, the symmetric and the antisymmetric spin parameter, respectively. The fiducial values of $t_c, \phi_c, \chi_s, \chi_a$ are set to be zero. We impose the priors on the parameters having domain of definition; $\log{\eta}$, $\phi_c$, angular parameters $(\theta_s, \phi_s, \cos{\iota}, \psi_p)$, and spin parameters $(\chi_s, \chi_a)$.

\begin{figure}
\begin{center}
\includegraphics[width=\hsize]{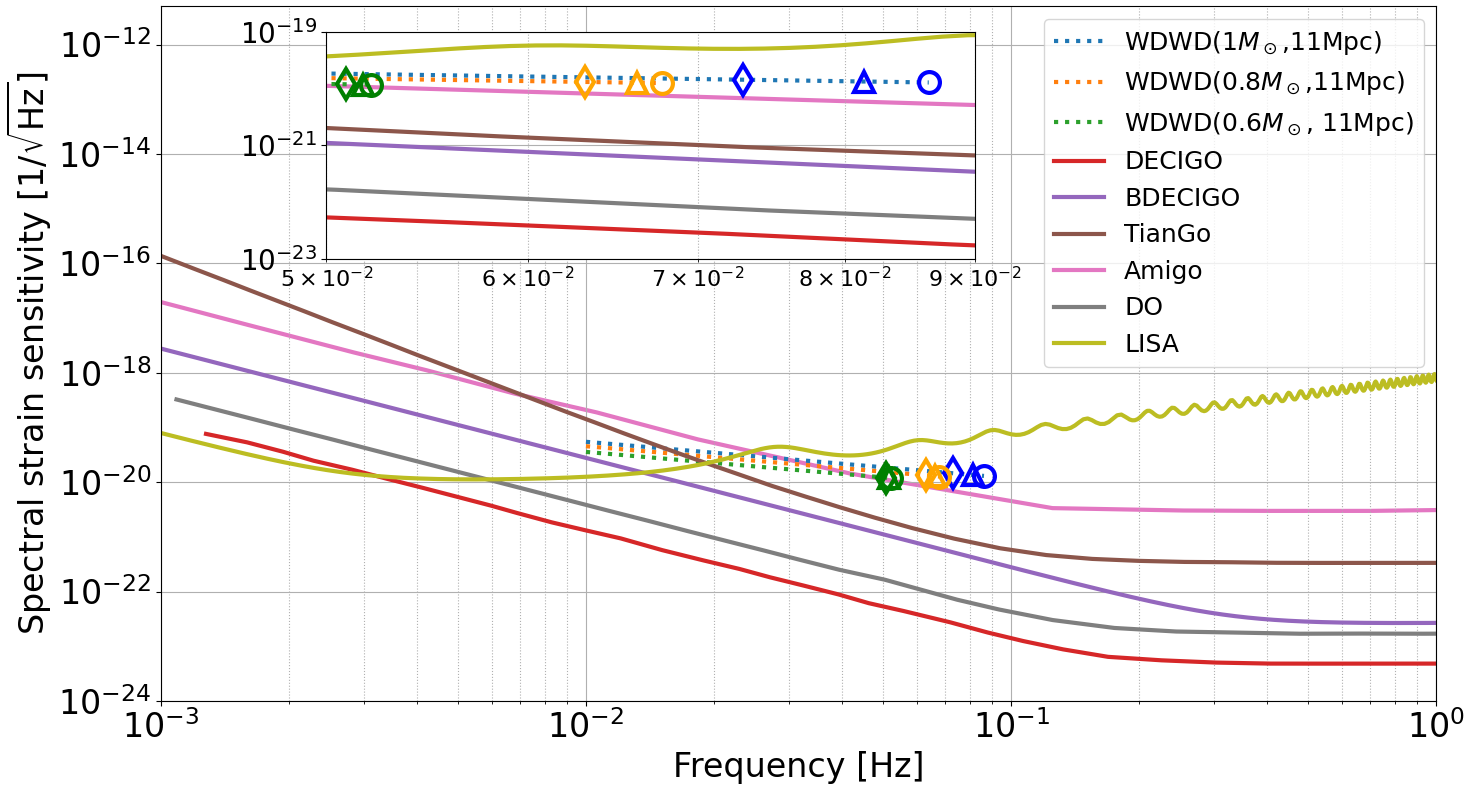}
\caption{ {Spectral strain sensitivities of the decihertz GW detectors and the equivalents for the source amplitudes of the inspiral GWs from equal-mass WD-WDs with different masses at $11\unit{Mpc}$.} The upper cutoff frequency is given by the frequency at the coalescence from \Eq{upper_frequency}. The diamonds, triangles, and circle points correspond to 10 years, 3 years before merger, and the upper cutoff frequency, respectively.}
\label{sensitivity}
\end{center}
\end{figure}

We set the upper cutoff frequency to the frequency at the coalescence time. Here, we roughly give the frequency at which the distance between the WDs is equal to the sum of the WD radius
\begin{equation}
f_{\rm{max}}\simeq\frac{1}{\pi}\sqrt{\frac{GM_{\rm tot}}{(R_1+R_2)^3}},
\label{upper_frequency}
\end{equation}
{from Kepler's 3rd law.}
The mass--radius relation for WDs is given by the equilibrium condition for the gravitational pressure and electron pressure \citep{Koester1990}.
For calculation of the upper cutoff frequency, we adopt the following mass--radius relation 
\begin{equation}
R_*\sim0.011\left(\frac{M_*}{M_\odot}\right)^{-1/3}R_\odot,
\label{R-M}
\end{equation}
based on the fitting of the observational data given by \cite{Magano2017}. Once the upper cutoff frequency is determined, the lower cutoff frequency is obtained by the integration time or the observational time using \Eq{tf}. {{\Fig{sensitivity} shows the inspiral GW strains from equal-mass WD-WDs having three different masses together with the spectral strain sensitivities of the decihertz GW telescopes. The plotted detector sensitivities are the root of the power spectral densities of their design sensitivities. The GW strains are plotted from an equivalent source amplitude of $2f^{1/2}|\tilde{h}(f)|$ \citep{Moore2015}.  The circle points denote the upper cutoff frequency, the diamond points correspond to 10 years before coalescence and the triangle points correspond to 3 years before coalescence, which are calculated from the time of merger and \Eq{tf}.} Compared to the GWs from compact binary mergers such as black holes and neutron stars observed at present by the ground-based detectors, the GWs from WD-WDs are almost monochromatic. However, we observe the slight frequency sweep effect in the decihertz band compared to that in LISA band.}

{We evaluated the estimation errors of the WD-WD parameters by Fisher analysis with DECIGO as the representative of decihertz gravitational wave detectors.} We assume that DECIGO is composed of three interferometers sharing the arms and has its design sensitivity. We also assume that its orbit is heliocentric orbit. The low-frequency approximation can be applied in all following calculations because the transfer frequency of the detector is $f_*:=c/(2\pi L)\sim 48 \unit{Hz}$ corresponding to the arm length of $L=1000 \unit{km}$. {Thereby, we ignore the transfer function in the antenna pattern functions \citep{Romano2017}. Unless otherwise noted, we conduct parameter estimation for the 100 binary systems whose distance is fixed to $11\unit{Mpc}$, within which about one type Ia SN event is expected per year from \Eq{11Mpc_rate}, to estimate the typical but conservative values of the expected errors. We mainly show the median values of the estimated errors for such multiple sources when the angular parameters $(\cos{\theta_s}, \phi_s, \cos{\iota}, \psi_p )$ are randomly distributed.}

\section{Results}
\label{Results}
\subsection{Full period observation}
\begin{figure}
 \begin{center}
 \includegraphics[width=\hsize]{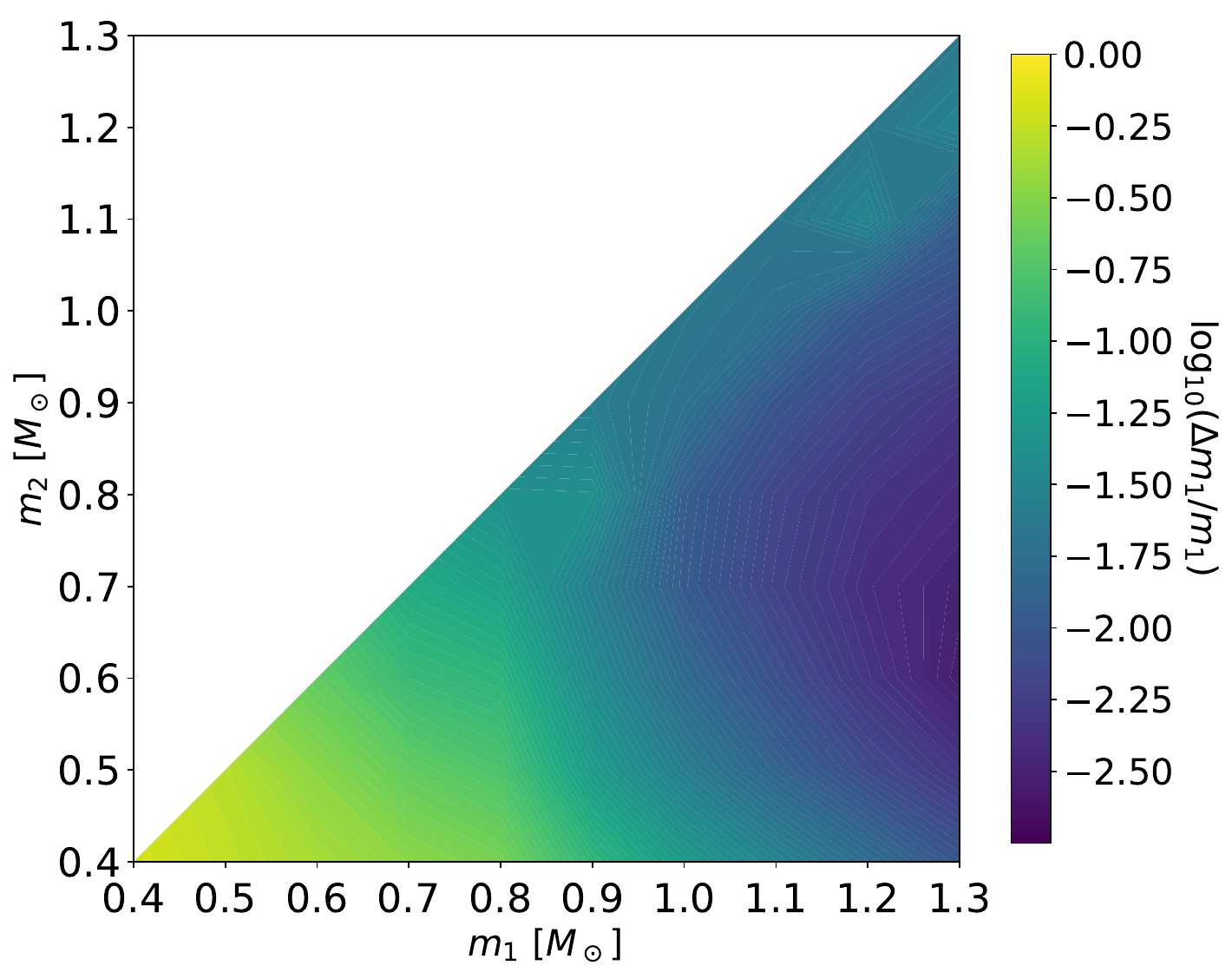}
 \end{center}
 \caption{Median vales of the logarithms of the fractional estimation errors of the primary WD component mass $\log_{10}(\Delta m_1/m_1)$ for $m_1-m_2$ WD-WDs at $11\unit{Mpc}$. {One Ia supernova event is expected per year within the distance from \Eq{11Mpc_rate}.} The full observations during the $3\unit{yr}$ operation period are assumed.}
 \label{m1_error}
\end{figure}

\begin{figure}
 \begin{center}
 \includegraphics[width=\hsize]{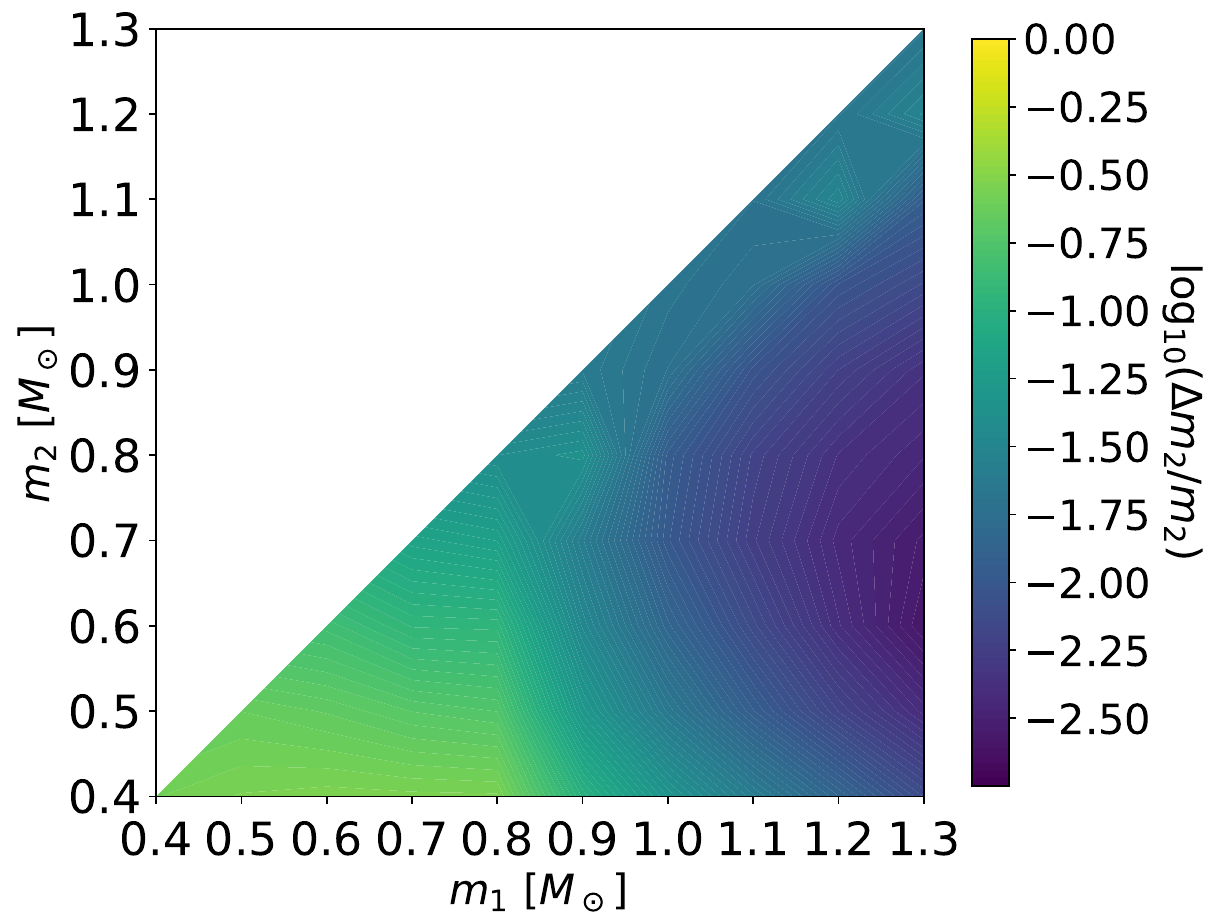}
 \end{center}
 \caption{Similar to \Fig{m1_error} but for the secondary WD component mass.}
 \label{m2_error}
\end{figure}

First, we consider full observations during the $3\unit{yr}$ operation period.  Figures. \ref{m1_error} and \ref{m2_error} show the fractional errors for the primary and the secondary WD component mass estimated by the Fisher analysis, respectively. The errors are estimated for 100 sources whose angular parameters are random for each mass combination by varying the component mass from $0.4 M_{\odot}$ to $1.3 M_{\odot}$ in $0.1 M_{\odot}$ intervals, and the median value are shown in the color maps. The minimum median value of SNR is 9.59 for the $0.4M_{\odot}-0.4M_{\odot}$ WD binaries and the maximum median value is 1240 for the $1.3M_{\odot}-1.3M_{\odot}$ WD binaries. The results show that the mass ratio can be measured, and then the WD component mass can be identified in most of the $m_1-m_2$ parameter space.
The errors of the secondary WDs are almost the same, slightly smaller than those of the primary WDs.
As the chirp mass is very well determined from the phase of the waveform, the error of the component mass is determined by the error of the mass ratio. If the mass of the primary WD is fixed, as the mass of the secondary WD decreases, the signal moves to a lower frequency band and the chirp effect decreases. As a result, the correlation among the chirp mass $\mathcal{M}$, the mass ratio $\eta$, the coalescence time $t_c$, and the right ascension $\phi_s$ increases, and the error of the mass ratio tends to increase.  Nevertheless, for the WD-WDs containing a WD with the mass heavier than $0.8M_{\odot}$, the masses were found to be determined with less than $\sim10\%$ precision even from the chirp effect alone. We found the best fractional error of $\sim 0.3\%$ for masses $1.3M_{\odot}-0.6M_{\odot}$. 

\begin{figure}
 \begin{center}
 \includegraphics[width=\hsize]{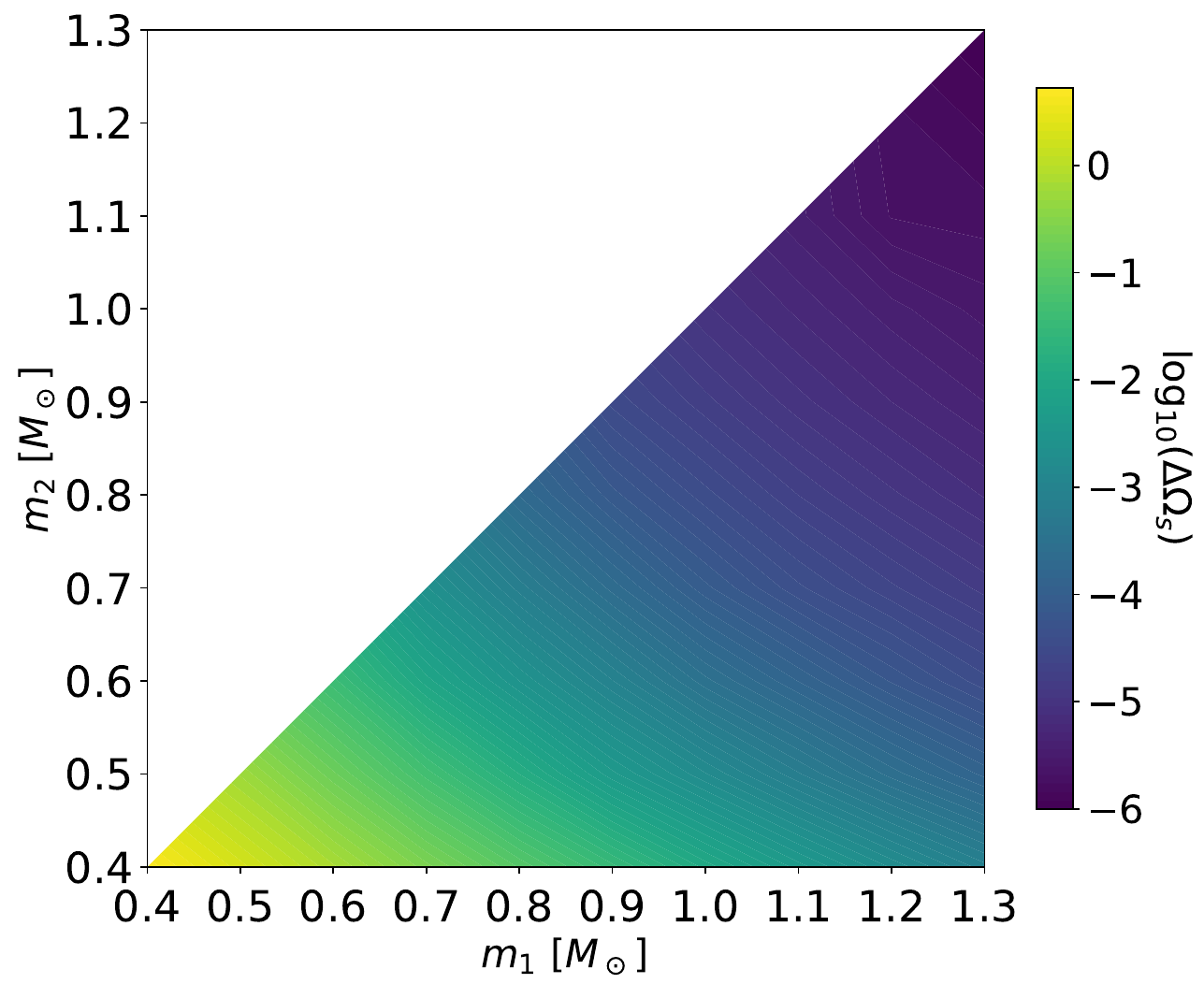}
 \caption{Median vales of the logarithms of the sky localization errors $\log_{10}(\Delta \Omega_s)$ for $m_1-m_2$ WD-WDs at $11\unit{Mpc}$. The full observations during the $3\unit{yr}$ operation period are assumed.}
 \label{omega_s_error}
 \end{center}
\end{figure}

\Fig{omega_s_error} shows the median values of the sky localization errors. {It indicate that the more massive the mass, the smaller the error.} The WD-WDs can be localized with the precision less than $\sim 5 \unit{deg^2}$ even for masses less than $0.8M_{\odot}$. The same trend as shown in \Fig{omega_s_error} was observed for $d_L$. We found that the luminosity distance can be determined with the precision less than $26\%$. Therefore, the WD-WDs can be localized with the 3D localization volume less than $\Delta V=d_{L}^3 \Delta \log{d_L}\Delta\Omega_s\sim0.5\unit{Mpc^3}$. In particular, for the WD-WDs containing a WD with the mass heavier than $0.8M_{\odot}$, we can determine the masses with the precision less than $\sim10\%$, and the position with the 3D localization volume less than $\sim1\times10^{-3}\unit{Mpc^3}$.

\subsection{For multi-messenger observation}
\begin{figure}
 \begin{center}
 \includegraphics[width=\hsize]{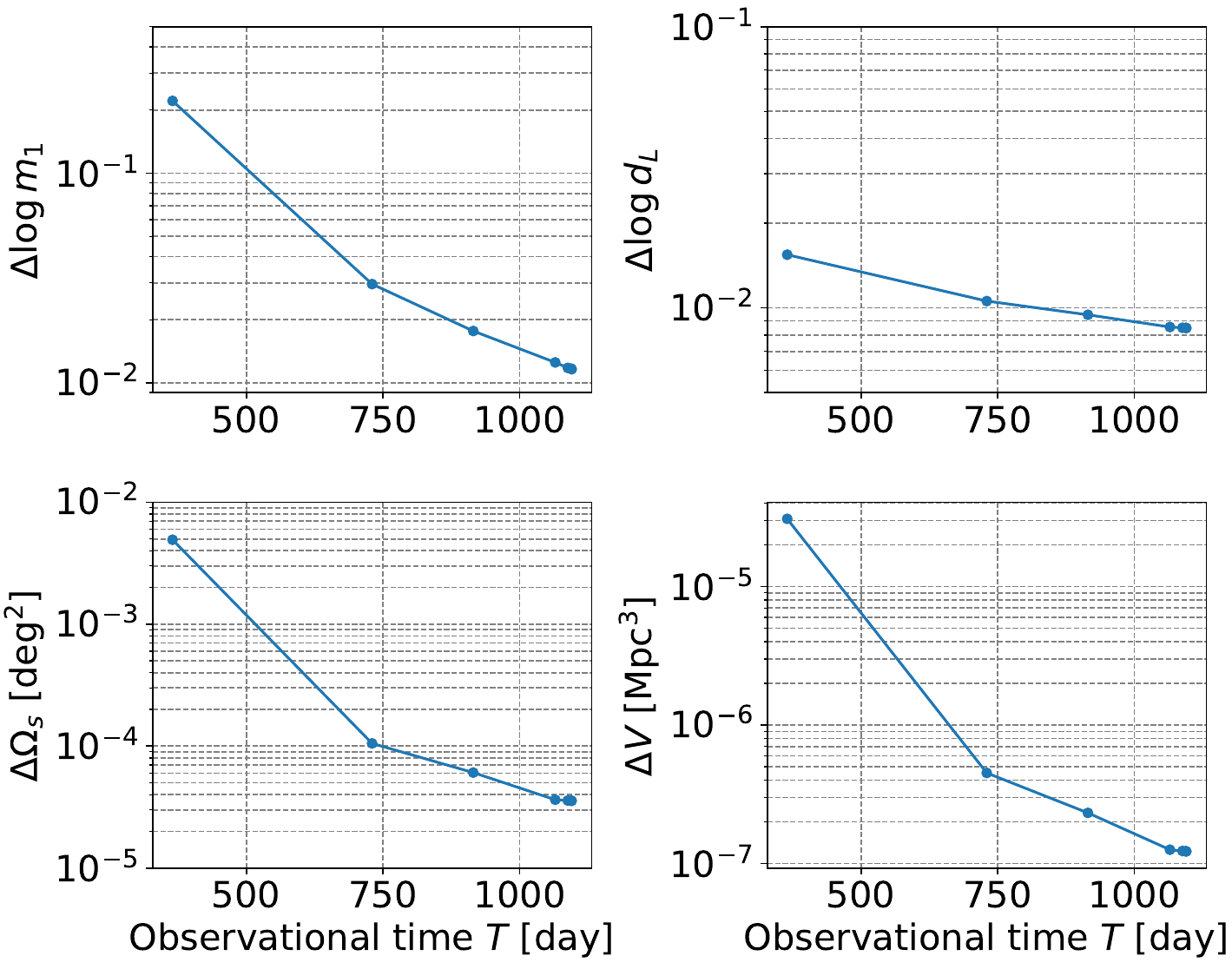}
 \caption{Estimation errors vs. observational time for $1.0M_{\odot}-0.8M_{\odot}$ WD-WDs at $11\unit{Mpc}$. The dots correspond to 2 years, 1 year, 6 months, 1 month, 1 week and 1 day before coalescence. The lower cutoff frequency is fixed to 3 years before the coalescence.}
 \label{error_multi_messenger}
 \end{center}
\end{figure}

Next, we analyze how the determination precision changes with the different observation time for multi-messenger observations. 
 {As one example, we consider the $1M_{\odot}-0.8M_{\odot}$ WD-WD case at $11\unit{Mpc}$. We change the upper cutoff frequency before coalescence and calculate the parameter estimation errors for 100 WD-WDs whose angular parameters are random, whereas the lower cutoff frequency is fixed to 3 years before coalescence. \Fig{error_multi_messenger} shows the dependence of the median value of the estimation errors on the observational time. The GWs emitted from WD-WDs are regarded as almost monochromatic waves. The SNR increases as ${\rm SNR}^2\sim {h_{\rm amp}T}/{S_n(f)}$. Thus, the errors such as $\Delta m_1$ and $\Delta d_L$ seem to improve in proportion to $T^{-1/2}$, but in comparison, the actual errors are worse in the short observation period. This is owing to the fact that the chirp effect becomes smaller for shorter observation periods, and the correlations among the mass ratio and other parameters mentioned above becomes larger, resulting in stronger parameter degeneracy. If we measure for at least two year, from three years to one year before the coalescence, we can determine in advance the masses of the WDs with the precision less than $10\%$ and its position with the 3D localization volume of $10^{-5}\unit{Mpc^3}$.\\}

{Deci-Hz space-based detectors such as DECIGO are expected to observe WD-WDs at {lager distances} than LISA.} To determine the typical decision precision for such distant WDs, we perform parameter estimation for WD-WDs fixed at a distance that gives the detection limit such that the SNR is approxomately $8$. Considering the $1M_{\odot}-0.8M_{\odot}$ WD-WD case again, we fix the redshift $z=0.08$ such that the median value of the SNR is approximately $8$. \Table{SNR8} shows the median values of the parameter estimation errors for the logarithms of the component masses, the luminosity distance, the sky localization, and the 3D localization volume.

\begin{table}
\caption{Medians of the parameter estimation errors for $1.0M_{\odot}-0.8M_{\odot}$ WD-WDs fixed at $z=0.08$ so that the median value of the SNR is nearly equal to 8.}
 \begin{center}
  \begin{tabular}{|c|c|} \hline
    parameter & WD-WD($1M_{\odot}$-$0.8M_{\odot}$, z=0.08) \\  \hline
	SNR &  8.21 \\ 
	$\Delta\ln{m_1}$ & $1.84\times10^{-1}$ \\
	$\Delta\ln{m_2}$ & $1.54\times10^{-1}$ \\
	$\Delta\ln{d_L}$    & $3.00\times10^{-1}$\\ 
	$\Delta\Omega_s[\rm{deg}^2]$ & $6.66\times10^{-2}$\\ 
	$\Delta V[\rm{Mpc^3}]$ & $3.18\times10^{2}$ \\ \hline
\end{tabular}
\label{SNR8}
 \end{center}
\end{table}

\section{Discussion \& Conclusions}
\label{Discussions}
 If the detection range of the deci-Hz detector is $D_L\sim\rm11~Mpc$, we may observe one WD-WD merger event per year.
We need $h<10^{-20}[\rm Hz^{-1/2}]$ as the detection sensitivity around ~0.1 Hz.
In \Fig{sensitivity}, the AMIGO's sensitivity is same as this value; hence, the SNR may be small $\sim1$--2.
Conversely, the sensitivities of TianGO and B-DECIGO are both $\sim3\times 10^{-22}[\rm Hz^{-1/2}]$, and 
DO's sensitivity is $\sim5\times10^{-23}[\rm Hz^{-1/2}]$.
Thus, TianGO, B-DECIGO, and DO are sufficiently sensitive to detect a WD-WD merger whose SNR is more than 8.

 {For example, in the case of $1M_{\odot}-0.8M_{\odot}$ WD-WD mergers, the median value of SNR is approximately equal to $8$ when $z=0.08\sim375\unit{Mpc}$.
The detectable volume can be estimated as $2.2\times10^8\unit{Mpc}^3$.
}
  {Using \Eq{rate}, it is expected that approximately 6600 Ia supernovae would occur per year during the observational period within the range, assuming that all Ia supernovae are caused by $1M_{\odot}-0.8M_{\odot}$ WD-WD mergers. }
 
Furthermore, DECIGO can detect many inspirals of WD-WDs more than 3 years before their mergers at the {0.01--0.1 Hz} range. Thus, we can investigate the mass and separation distribution of the WD-WDs and the relation between them and their host galaxies. Note that the detection rate estimate strongly depends on the masses of WDs, as the GW frequency is proportional to the total mass of WD-WD (Eqs.\,\ref{upper_frequency} and \,\ref{R-M}). 

 {The 3D localization volume of DECIGO $ d_L^3\Delta \ln d_L\Delta\Omega_s$ for $1M_{\odot}-0.8M_{\odot}$ WD-WD mergers at $z=0.08$ is $318~\rm Mpc^3$.
Conversely,the 3D localization volume WD-WD merger within $d_L\sim 11 \rm~Mpc~$ is $\sim 10^{-7}\rm~Mpc^3$ (\Fig{error_multi_messenger}). 
{DECIGO provides a more useful and easier way to determine the host galaxy and its location compared to the host galaxy identification by LIGO observation.}
This value is proportional to $d_L^{6}$ due to $\Delta \ln d_L \propto d_L$ and $\Delta \Omega_s \propto d_L^2$.
Thus, the 3D localization volume for $1M_{\odot}-0.8M_{\odot}$ WD-WD mergers within $d_L\sim300 \rm~Mpc~$$ (z=0.065)$ is $\sim100\rm~Mpc^3$.
The Milky Way-like galaxy density is one galaxy per 100 Mpc$^3$ \citep{Kopparapu2008}; hence, we can identify host galaxies for many WD-WD mergers using the GW detection only. }
{It is useful and much easier to do follow-up observations and to identify the EM counterpart than the case of LIGO observation.}

\begin{figure}
  \begin{center}
    \includegraphics[width=\hsize]{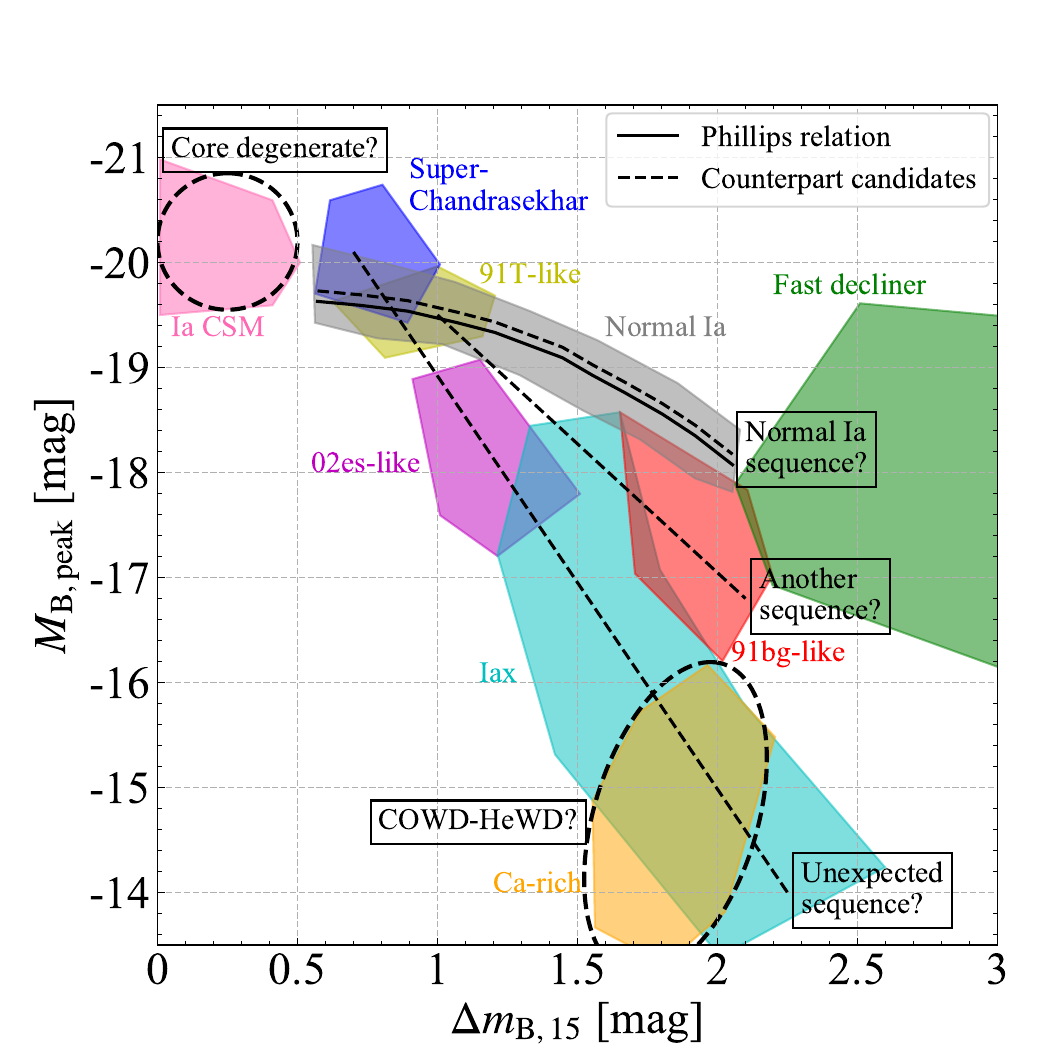}
  \end{center}
  \caption{Peak magnitudes and decline rates in B band for normal Ia
    supernovae and subclasses of Ia supernovae, which is reproduced from figure 1 in \cite{Taubenberger2017}. The solid black
    curve shows the Phillips relation. The dashed black curves
    indicate counterpart candidates of WD-WD mergers supposed in the
    main text.}
  \label{fig:ia-diagram}
\end{figure}

 {Multi-messenger (GW and EM) observations, will
  enable us to put unprecedented constraints on WD-WD merger outcomes
  and possibly Ia progenitors. First, we can confirm if WD-WD
  mergers cause prompt explosions and accompany bright astronomical
  transients as summarized in Figure \ref{fig:ia-diagram}. Then, if WD-WD
  mergers result in any transients, we can attribute  to
  some types of thermonuclear transients. It is given that WD
  thermonuclear explosions are involved in various types of
  transients: normal Ia supernovae, subclasses of Ia supernovae
  (e.g. type Ia supernovae associated with circum-stellar matter
  (CSM), Super-Chandrasekhar Ia supernovae, SN 1991T-likes, SN
  1991bg-likes, SN 2002es-likes, and Iax supernovae), and Ca-rich
  transients \cite[see review by][]{Jha2019}.}

 {Owing to the high-quality mass estimate by GW observations (see Figures \ref{m1_error} and \ref{m2_error}), we can assess the importance of WD masses. For example, let us assume that normal Ia supernovae results from WD-WD mergers; they will change their peak magnitudes and decline rates along with the Phillips relation \citep{Phillips1993} with changing WD masses. Then, we can attribute the physical background of the Phillips relation to WD masses (see ``Normal Ia sequence'' in Figure \ref{fig:ia-diagram}). By means of numerical simulations of WD explosions, {\cite{Ruiter2013} indicated that the peak magnitudes depend on exploding WD masses, and} \cite{Shen2018} showed that the peak magnitudes and decline rates depend on exploding WD masses {partly} along with the Phillips relation. 
 Conversely, GW observations will provide information of WD masses independently of such numerical simulations, and can be combined with EM observational results of the peak magnitudes and decline rates. }

 {We further give three possibilities as examples. First,
  exploding WD masses may be responsible for the appearance of
  transients: normal Ia or subclasses of Ia supernovae (see ``another
  sequence'' in Figure \ref{fig:ia-diagram}). The violent merger
  model, one of the WD explosion models that may represent events after WD-WD mergers, is
  suggested to cause SN 1991-bg-likes for primary WD masses with $\sim
  0.9 M_\odot$ \citep{Pakmor2010}, and normal Ia supernovae for
  primary WD masses with $\sim 1.1 M_\odot$
  \citep{Pakmor2012}. Second, the difference between exploding WD masses
  may yield an ``unexpected sequence'' (see Figure
  \ref{fig:ia-diagram}), ranging over Super-Chandrasekhar, 91T-likes,
  02es-likes, and Iax supernovae. We should note that some of them may
  have {\it SD progenitors}, but not DD progenitors. Some
  super-Chandrasekhar Ia supernovae indicate massive CSM like SN
  2012dn \citep{Yamanaka2016}. Iax supernovae can have bright
  companion stars like SN 2012Z \citep{McCully2014}. This is why
  we call it ``unexpected'' sequence.  Third, a WD-WD merger, one of
  which has a small mass ($\lesssim 0.5M_\odot$), can generate a
  Ca-rich transient \citep{Perets2010} as seen in ``COWD-HeWD'' in
  Figure \ref{fig:ia-diagram}. Note that a $\lesssim 0.5M_\odot$ WD is
  thought as a helium (He) WD.}

 {The direct measurement of WD masses will also facilitate the determination of the combustion process of WDs, and emission process of Ia supernova ejecta. As for the combustion process, the peak magnitudes can be converted into radioactive nuclear masses, $^{56}$Ni masses {\citep{Ruiter2013, Shen2018}.} Thus, we can assess if carbon detonation, a promising combustion process in sub-Chandrasekhar mass WDs, can yield EM-observed $^{56}$Ni masses from GW-observed WD masses. As for the emission process, the decline rates can be related to the opacity of supernova ejecta, WD masses {\citep{Hoeflich1996, Nugent1997, Maeda2003, Kasen2009}.} We can also verify if EM-observed decline rates are consistent with GW-observed WD masses.}

\begin{figure}
  \begin{center}
    \includegraphics[width=\hsize]{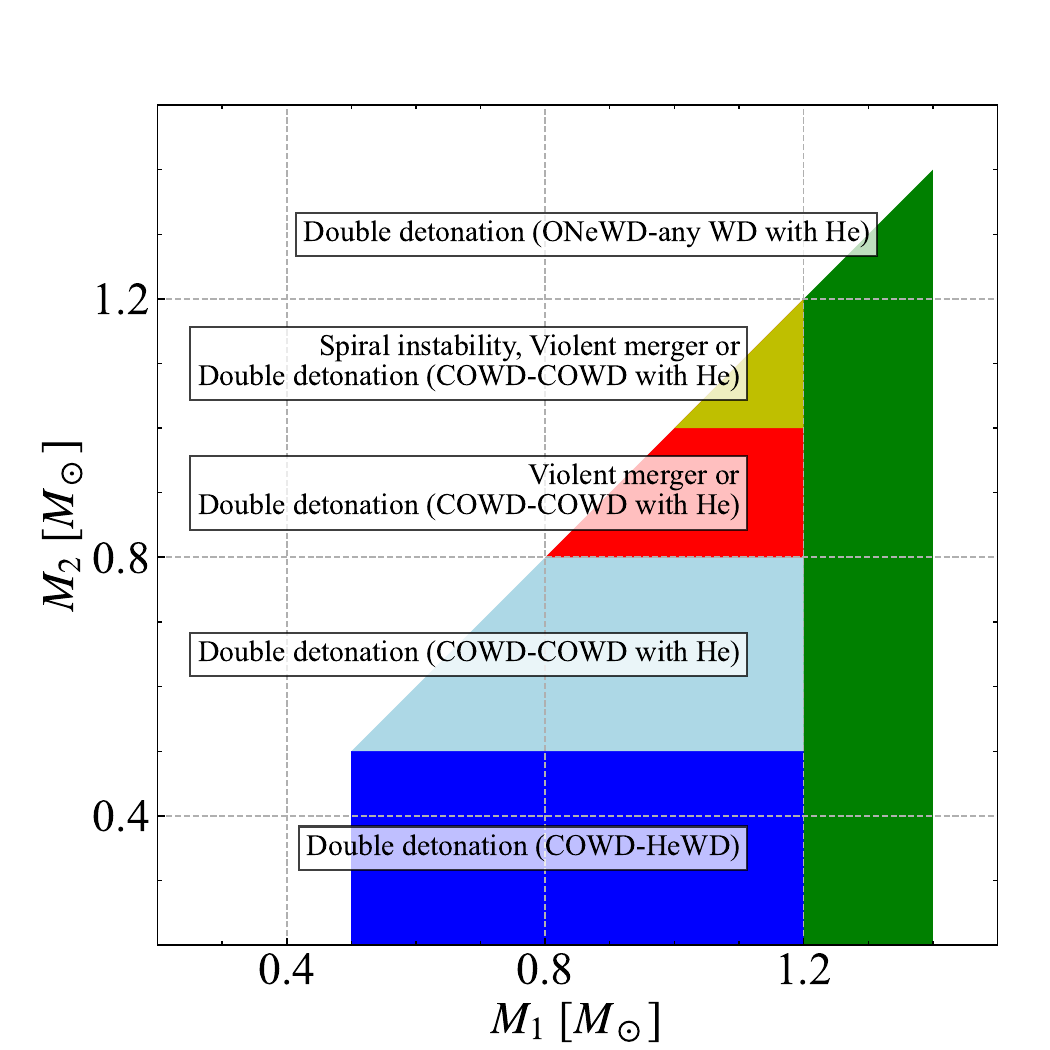}
  \end{center}
  \caption{Mass combination of WD-WD mergers, and candidates of
    ignition processes. He and C-O WDs are assumed to have
    $<0.5M_\odot$ and $0.5$--$1.2M_\odot$. C-O WDs are allowed to have
    He shells on their surfaces. The primary WDs are limited to C-O
    WDs. {The double detonation with any WD combinations in DD systems has been suggested by \cite{Guillochon2010, Fink2010, Woosley2011, Pakmor2013, Pakmor2021, Pakmor2022, Shen2014}, and \cite{Tanikawa2018, Tanikawa2019}. The (carbon-ignited) violent merger is numerically demonstrated by \cite{Pakmor2010, Pakmor2012} and \cite{Tanikawa2015}. The spiral instability is reported by \cite{Kashyap2015, Kashyap2017}. Double detonation in a DD system with an ONe WD can also occur according to \cite{Marquardt2015}.}}
  \label{fig:ignition-mass}
\end{figure}

 {Even if a WD-WD merger is a progenitor of any transient, the
  current GW analysis may not be sufficient to identify the ignition process responsible for the transient as seen in Figure \ref{fig:ignition-mass}. Several ignition processes in WD-WD mergers have been suggested: the double detonation in a mass transfer phase like the D$^6$ model \citep{Guillochon2010, Fink2010, Woosley2011, Pakmor2013, Pakmor2021, Pakmor2022, Shen2014, Marquardt2015, Tanikawa2018, Tanikawa2019}, the carbon detonation in a merger phase like the violent merger model \citep{Pakmor2010, Pakmor2012, Tanikawa2015}, and the carbon detonation via spiral instability in an early and asymmetric accretion disk phase \citep{Kashyap2015, Kashyap2017}. If the lighter WD has a sufficiently small mass (say $\lesssim 0.8M_\odot$), we can reject the violent merger and spiral instability, as both models need massive secondary WDs, $\gtrsim 0.8M_\odot$ and $\gtrsim 1.0M_\odot$, respectively
  \citep[][respectively]{Sato2016, Kashyap2017}. {If the heavier WD has a sufficiently large mass (say $\gtrsim 1.2M_\odot$), we can identify the exploding WD as an oxygen-neon (ONe) WD and  adopt the model of \cite{Marquardt2015}.} However, if the secondary WD mass is close to $1.0M_\odot$, we cannot reject any ignition processes, because all the ignition processes are possible in such systems. Note that we may support (or reject) the double detonation model, if we confirm the presence (or absence) of He-detonation ashes by means of detailed spectroscopic observations like MUSSES1604D \citep{Jiang2017} and ZTF18aaqeas/SN 2018byg \citep{De2019}.}

\begin{figure}
  \begin{center}
    \includegraphics[width=\hsize]{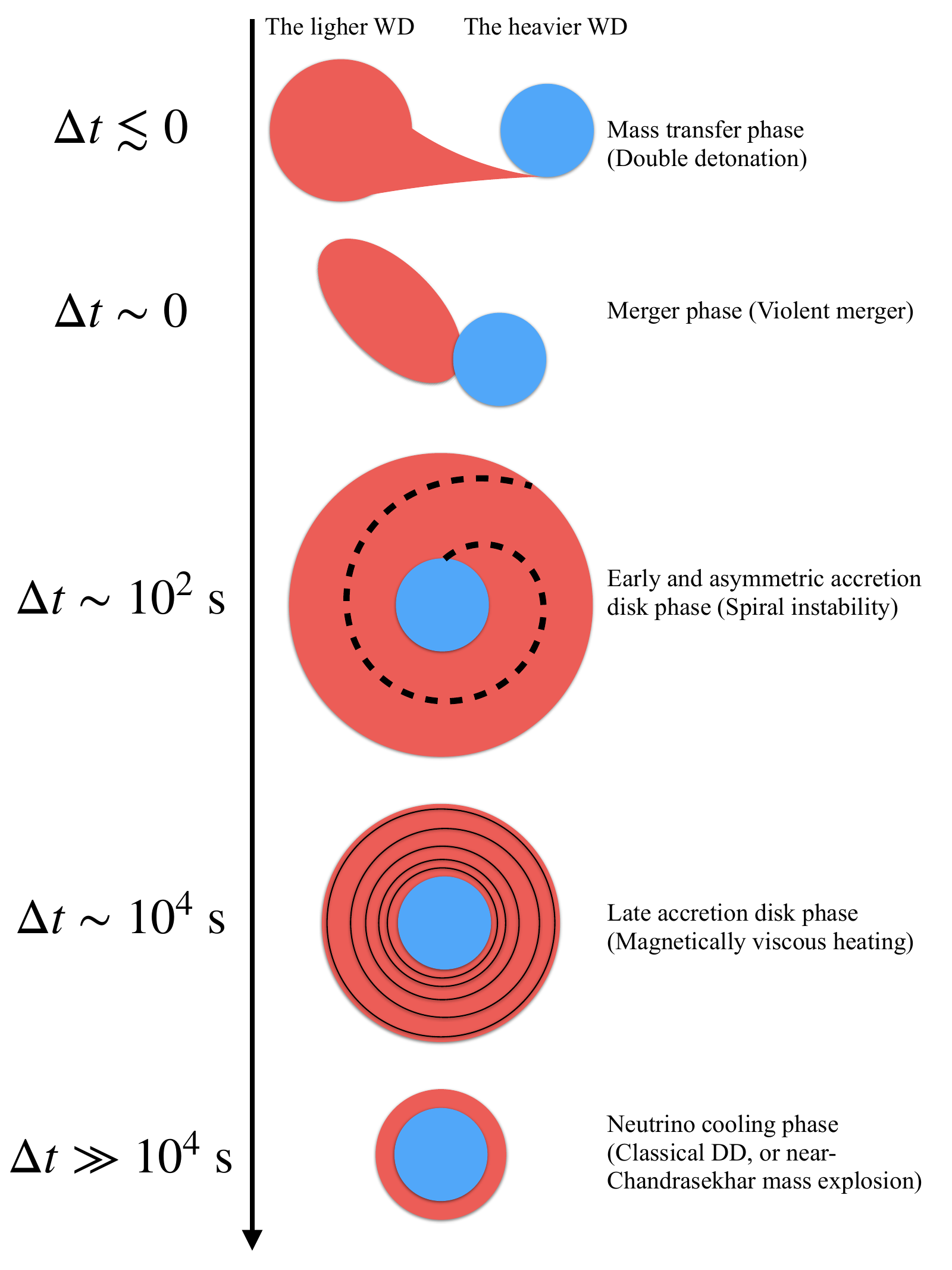}
  \end{center}
  \caption{Schematic of the relation between ignition processes and WD-WD phases, where $\Delta t$ is defined as the difference between the ignition time and merger time. Note that, for negative $\Delta t$, the merger time would be given if the ignition did not occur. Double detonation can occur before WD-WD mergers. The violent merger initiates at the moment of the merger. Spiral instability can ignite explosive carbon burning at an early and asymmetric accretion disk phase, where the accretion disk is made of the lighter WD tidally disrupted. In a late accretion disk phase, magnetic viscosity heats the system, although there is no scenario for WD explosion in this phase. In a neutrino cooling phase, classical DD scenario (or near-Chandrasekhar mass explosion) may work. {The double detonation, violent merger, and spiral instability are studied in references cited in the caption of Figure \ref{fig:ignition-mass}. \cite{Dan2015} investigated WD explosion at a similar phase to the phase of magnetically viscous heating, although they artificially put a hot spot to start carbon detonation . The classical DD was suggested by \cite{Webbink1984} and \cite{Iben1984}.}}
  \label{fig:ignition-time}
\end{figure}

 {In our subsequent paper (Takeda et al. in prep.), we will analyze WD-WD mergers in more detail, and distinguish GW signals of ignition processes (see Figure \ref{fig:ignition-time}). The aforementioned ignition processes should have different GW signals, because they cause explosions at different times with respect to the WD-WD merger time ($\Delta t$). WDs explode long before their merger or in the mass transfer phase ($\Delta t \lesssim 0$) in the double detonation
  models, nearly at the moment of their merger ($\Delta t \sim 0$) in the violent merger model, and several $100$ seconds after their merger or just after formation of an accretion disk made from a tidally disrupted WD ($\Delta t \sim 100$ seconds) in the spiral instability model. The signals from GWs just before they disappear due to explosions or mergers may be different among these processes.}

 {Multi-messenger (GW and EM) observations, will be
  helpful again to specify alternative ignition processes. GW
  observations can determine the time of disappearance of GW signals
  with high accuracy. EM observations can give the ignition time with
  accuracy of $\sim 1$ hour even in the present day, where the Zwicky
  Transient Facility archives 2-hour cadence \citep{Bellm2014}, for
  example. Thus, we can specify ignition processes in which the WD
  explosion is later than a WD-WD merger by $\gtrsim 1$ hour. If an
  EM-observed WD explosion happens at $10^4$ seconds after a GW-observed
  WD-WD merger ($\Delta t \sim 10^4$ seconds), magnetically viscous
  heating in the accretion disk may contribute to the ignition. Note
  that there is no scenario for WD explosions in this phase to our
  knowledge. For the case of $\Delta t \gg 10^4$ seconds, we may
  conclude that the WD explosion occurs along with the classical DD
  scenario, or near-Chandrasekhar mass explosion through neutrino
  cooling \citep[e.g.][]{Webbink1984,Iben1984}. 
  If that is true, the WD-WD merger remnant experiences slow mergeing \citep{Yoon2007},
  avoiding quasi-static carbon burning, which converts the remnants
  into oxygen-neon-magnesium WDs \citep{Saio1985, Schwab2016}. Even if
  we do not have any transients, we cannot rule out the possibility of a
  near-Chandrasekhar mass explosion. The explosion can have a delay time
  of $\sim 10^5$ yr from the WD-WD merger \citep{Yoon2007}, much more
  than a human (or civilization) lifespan.}

\section*{Acknowledgements}
We would like to greatly thank the anonymous referee for their useful comments to improve our paper.
We would like to thank Makoto G. Ando, Takashi Nakamura, Atsushi Nishizawa, Masaki Ando, Kazuhiro Shimasaku, Naoki Seto, Takahiro Sudoh for useful discussion.
This work was supported by JSPS KAKENHI Grant No. 17H06360 (AT), 18J00558(TK), 18J21016(HT), 19H00704(HY), 19K03907(AT), and 21K13915(TK).
TK acknowledges support from the University of Tokyo Young Excellent Researcher program.
HT acknowledge financial support received from the Advanced Leading Graduate Course for Photon Science
(ALPS) program at the University of Tokyo.
We thank Editage (www.editage.com) for English language editing.

\end{document}